\documentclass[twocolumn,aps,preprintnumbers,pra,showpacs,amsfonts,amsmath,amssymb,superscriptaddress,floatfix]
{revtex4}
\usepackage[dvips]{graphicx}
\newcommand{\affA}{%
     Quantum Information Technology Group, 
     National Institute of Information and Communications Technology
     (NICT), \\
     4-2-1 Nukui-kitamachi, Koganei, Tokyo 184-8795, Japan}
\newcommand{\affB}{%
     CREST, Japan Science and Technology Agency, 
     1-9-9 Yaesu, Chuoh-ku, Tokyo 103-0028, Japan}
\newcommand{\affC}{%
     Quantum Information Theory Group, Institute of Theoretical Physics, \\ 
     Universit\"{a}t Erlangen-N\"{u}rnberg, 91058 Erlangen, Germany}
\newcommand{\affD}{%
     Institute for Quantum Computing, 
     University of Waterloo, \\ 
     200 University Avenue West, Waterloo, Ontario, N2L 3G1, Canada}

\begin{document}
\title{Binary projective measurement via linear optics and 
photon counting}
\date{\today}
\author{Masahiro Takeoka}
\author{Masahide Sasaki}%
\address{\affA}%
\address{\affB}%

\author{Norbert L\"{u}tkenhaus}%
\address{\affC}%
\address{\affD}%

\begin{abstract}

We investigate the implementation of binary projective measurements with linear optics. 
This problem can be viewed  as a single-shot discrimination of two orthogonal 
pure quantum states. We show that any two orthogonal states 
can be perfectly discriminated using only linear optics, photon counting, 
coherent ancillary states, and feedforward. The statement holds
in the asymptotic limit of large number of these physical resources. 

\end{abstract}

\pacs{03.67.Hk, 03.65.Ta, 42.50.Dv}

\maketitle

Projection measurements play an essential role 
in photonic quantum-information protocols. 
In these applications, generally, 
a projection onto superposition states or entangled states of optical fields 
is required. Physically, 
it is a highly nontrivial problem how to implement such a measurement.

One plausible approach is to use linear optics and 
classical feedforward associated with a partial measurement. 
For example, a universal quantum computation scheme 
for photonic-qubit states has been proposed, which utilizes 
only linear optics, photon counting, and 
highly entangled auxiliary states of $n$ photons generated 
by probabilistic gate operations \cite{KLM01}. 
In principle, 
it works with unit success probability in the asymptotic 
limit of large $n$.  
It is, however, still a nontrivial question how to prepare entangled 
ancillae even for modest $n$.

In this paper, we discuss the linear optics implementation 
of a measurement which effects a projection onto 
two orthogonal states $\{|\Psi\rangle, |\Phi\rangle\}$. 
This is equivalent to the problem of discriminating 
two orthogonal quantum signals $\{|\Psi\rangle,|\Phi\rangle\}$
unambiguously \cite{vanLoock03,comment1}. 
We show that, in the asymptotic limit of a large number 
of partial measurements, 
one can perfectly discriminate the two states with linear optics, 
photon counting, and feedforward, but {\it without} any non-classical 
auxiliary states. 
Even in the worst case, the average error probability 
of discrimination approaches zero 
with the scaling factor of $N^{-1/3}$ 
where $N$ is the number of the partial measurements.
Note that the signal space is two-dimensional but $|\Psi\rangle$ and 
$|\Phi\rangle$ can be any physical states defined in a larger space, 
e.g. qubit states, continuous variable states, etc.

Before discussing a linear optics implementation, 
it is worth mentioning a result concerning the distinguishability of 
two orthogonal multi-partite
states via local operations 
and classical communication (LOCC). 
The necessary condition for exact local distinguishability is that, 
after doing a measurement at some local site, every possible remaining states 
must be orthogonal to each other. 
Walgate {\it et al.} \cite{Walgate00} showed that 
there always exists a local projective measurement 
satisfying this orthogonality condition for any set of two orthogonal states. 
Thus one can perfectly discriminate them via 
a series of local projective measurements where the choice 
of the measurement basis at each local site is conditioned 
on the previous measurement outcomes. 
This result means that if one can show a physical scheme that can exactly 
discriminate any two orthogonal {\it single-mode} states, 
its sequential application can achieve an exact discrimination of 
any two orthogonal {\it multi-mode} states. 
In the following, therefore, we concentrate on a discrimination of 
two single-mode states.

\begin{figure}
\begin{center}
\includegraphics[width=0.9\linewidth]{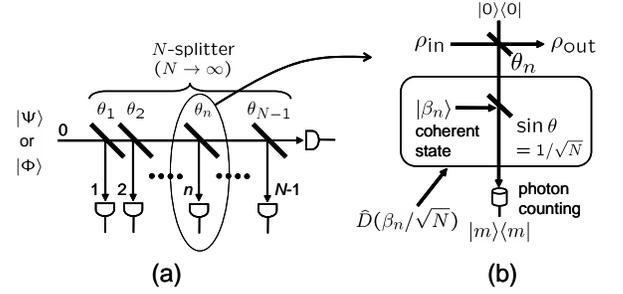}   %
\caption{\label{fig:schematic}
(a) $N$-splitter, and (b) a measurement apparatus at each step.
A displacement operation $\hat{D}(\beta_i/\sqrt{N})$ 
is realized by combining the signal with 
a coherent state local oscillator $|\beta_i/\sqrt{N\sin\theta}\rangle$ 
via a beamsplitter with sufficiently small power reflectance of 
$\sin^2\theta$.
}
\end{center}
\end{figure}

An arbitrary set of two orthogonal single-mode states are described by 
\begin{eqnarray}
\label{eq:two_states}
|\Psi\rangle = \sum^{\infty}_{m=0} c_m |m\rangle_0 , \qquad
|\Phi\rangle = \sum^{\infty}_{m=0} d_m |m\rangle_0 , 
\end{eqnarray}
where $|m\rangle$ is an $m$-photon number state and 
$\langle\Psi|\Phi\rangle = 
\sum^{\infty}_{m=0} c_m^* d_m = 0$. 
Figure~\ref{fig:schematic} is the schematic of the measurement apparatus. 
The states are equally split into $N$ modes by  $N-1$ asymmetric beamsplitters 
\cite{vanLoock00},
\begin{eqnarray}
\label{eq:N-splitter}
& & \hat{B}_{N-1,0} (\theta_{N-1}) \hat{B}_{N-2,0} (\theta_{N-2})
\cdots \hat{B}_{1,0} (\theta_{1}) |0\rangle^{\otimes N-1} |\Psi\rangle_0 
\nonumber\\ & & 
= e^{ - \hat{a}_{N-1}^{\dagger} \hat{a}_0 } 
\cdots 
e^{ - \hat{a}_{1}^{\dagger} \hat{a}_0 } 
e^{ \hat{a}_0^{\dagger} \hat{a}_0 \ln \left( 1/\sqrt{N} \right) }
|0\rangle^{\otimes N-1} |\Psi\rangle_0 
\nonumber\\ & & 
\equiv \hat{N}_{BS} |\Psi\rangle_0 ,
\end{eqnarray}
where 
$\hat{B}_{i,0} (\theta_i) = \exp [ \theta_i(\hat{a}^{\dagger}_i \hat{a}_0 
- \hat{a}_i \hat{a}^{\dagger}_0 ) ]$
\cite{Barnett97} and 
$\tan\theta_i = 1/\sqrt{N-i}$.
The input is symmetrically split to $N$ modes with 
the effective power reflectance of $1/N$. 
Then, at each output port, one makes some measurement 
by using linear optics and photon counters, where the information about 
the measurement outcome is fed forward to design 
the next measurement. 
It should be noted that this is a generalized version of the scheme so-called 
``Dolinar receiver'' \cite{Dolinar73,Geremia04,Takeoka05}
which was originally proposed as a physical model attaining 
the minimum error discrimination of the binary coherent signals
$\{|\alpha\rangle, |-\alpha\rangle\}$.

We briefly sketch how two states are discriminated 
by such a scheme in the limit of $N\to\infty$ and then 
provide a rigorous proof. 
Suppose one inserts $|\Psi\rangle$ or $|\Phi\rangle$ 
into the first beamsplitter. 
For sufficiently small $1/N$, 
the reflectance of multi-photons can be neglected. 
The states after beamsplitting are approximated to be
$\hat{B}_{1,0}(\theta_1) |0\rangle_1 |\Psi\rangle_0
\approx |0\rangle_1 |\eta_0\rangle_0 + 
N^{-1/2} |1\rangle_1 |\eta_1\rangle_0$, and 
$\hat{B}_{1,0}(\theta_1) |0\rangle_1 |\Phi\rangle_0
\approx |0\rangle_1 |\nu_0\rangle_0 + 
N^{-1/2} |1\rangle_1 |\nu_1\rangle_0$, 
where, 
$\langle \eta_0 | \nu_0 \rangle + \langle \eta_1 | \nu_1 \rangle 
/N \approx 0 , $
since a beamsplitting operation is unitary. 
Then mode 1 is measured. 
The measurement here is required 
to maintain the orthogonality of any conditional outputs of 
$|\Psi\rangle$ and $|\Phi\rangle$. 
The local measurement satisfying this condition is 
described by a two-dimensional projective measurement, 
\begin{eqnarray}
\label{eq:approx_projection_vectors}
|\pi_0\rangle & = & 
\mathcal{N}_{p0} \left\{  |0\rangle 
+ \frac{1}{X^*}\left(1-\sqrt{1+|X|^2}\right) |1\rangle \right\} 
\nonumber\\ & = & 
\mathcal{N}_{p0} \left\{ |0\rangle 
- (X + O(X^2)) |1\rangle \right\}, \\
\label{eq:approx_projection_vectors'}
|\pi_1\rangle & = & 
\mathcal{N}_{p1} \left\{ (X^* + O(X^2)) |0\rangle 
+ |1\rangle \right\}. 
\end{eqnarray}
where, $\mathcal{N}_{p0}$ and $\mathcal{N}_{p1}$ are 
the normalization factors and 
\begin{equation}
\label{eq:X}
X = \frac{ 2( \langle\nu_0|\eta_1\rangle \langle\eta_1|\nu_1\rangle
- \langle\eta_0|\nu_1\rangle \langle\nu_1|\eta_1\rangle ) 
}{ \sqrt{N}
(|\langle\eta_0|\nu_1\rangle|^2 - |\langle\eta_1|\nu_0\rangle|^2) }. 
\end{equation}
Here, we have assumed $|\langle\eta_0|\nu_1\rangle|^2 
- |\langle\eta_1|\nu_0\rangle|^2 \neq 0$ which implies 
$X \propto 1/\sqrt{N}$ and thus we can take $|X| \ll 1$ 
in the limit of large $N$. 
The other case, i.e. $|\langle\eta_0|\nu_1\rangle|^2 
- |\langle\eta_1|\nu_0\rangle|^2 = 0$, will be discussed later.
Under this assumption, the projective measurement 
of Eqs.~(\ref{eq:approx_projection_vectors}) 
and (\ref{eq:approx_projection_vectors'}) can be implemented 
by the displacement operation $\hat{D}(\beta_1/\sqrt{N})$ 
and photon counting as shown in Fig.~\ref{fig:schematic}(b). 
Since both the signal and displacement are sufficiently weak, 
the corresponding measurement vectors are described by 
\begin{eqnarray}
\label{eq:approx_displaced_pc_0}
\hat{D}^{\dagger}
\left(\frac{\beta_1}{\sqrt{N}}\right) |0\rangle & \approx & 
e^{-|\beta_1|^2 /2N} 
\left( |0\rangle - \frac{\beta_1}{\sqrt{N}} |1\rangle \right), \\
\label{eq:approx_displaced_pc_1}
\hat{D}^{\dagger}
\left(\frac{\beta_1}{\sqrt{N}}\right) |1\rangle 
& \approx & 
e^{-|\beta_1|^2 /2N}
\left( \frac{\beta_1^*}{\sqrt{N}} |0\rangle + |1\rangle \right), 
\end{eqnarray}
which can be same as Eqs.~(\ref{eq:approx_projection_vectors}) 
and (\ref{eq:approx_projection_vectors'}) by choosing 
appropriate $\beta_1$.

The conditional states after the first measurement can be 
rewritten again as 
$|\Psi'\rangle = \sum^{\infty}_{m=0} c'_m |m\rangle$ and 
$|\Phi'\rangle = \sum^{\infty}_{m=0} d'_m |m\rangle$. 
Since $\hat{N}_{BS}$ splits a state symmetrically, 
one can repeat the same procedure for the remaining state
with the second beamsplitter, the displacement operation 
$\hat{D}(\beta_2/\sqrt{N})$, where $\beta_2$ is conditioned 
on the previous measurement outcome, and a photon counter. 
After repeating the same procedure to modes 1 to $N-1$ 
with appropriate $\beta_i$'s, the final states at mode 0 contain 
with dominating weight at most one photon and are still orthogonal 
to each other. 
As a consequence, applying the final ($N$-th) displacement 
and photon counting, one can exactly discriminate 
$|\Psi\rangle$ and $|\Phi\rangle$ with unit success probability.

Now, we discuss the scheme  rigorously, i.e. 
include the effects due to the multi-photon reflections 
at each beamsplitter, which contribute to 
the failure of the measurement or giving the incorrect decisions. 
Here, the input states $|\Psi\rangle$ and $|\Phi\rangle$ are 
always physical, that is, 
the average power of them are finite. 
Moreover, we assume that the probability distribution in photon number 
of those states decreases exponentially 
as $c_m \equiv \tilde{c}_m e^{-mx/2}$ 
where $x$ is a real positive number. 
The prior probabilities can be set to be equal
without loss of generality. 
Finally we assume that the average powers of 
local oscillators always satisfy 
$|\beta_i|^2 \le |C_{\beta_i}|^2 + O(1/N)$ 
where $C_{\beta_i}$ is a complex constant independent of $N$.

After finishing a whole process of $N$ measurement steps, 
one can classify the results according to 
the sequential patterns of detected photon numbers. 
Let us denote the events in which all the photon counters detect zero or 
one photon by `success' events and the others by `failure' events. 
Because of the symmetry of the $N$-beamsplitting, 
the probability of detecting $k$ photons at the $i$-th measurement 
{\it on average over all possible measurement patterns} is given by 
\cite{Barnett97}
\begin{eqnarray}
\label{eq:P_k_average}
P_k^{(i)} & = & \left| 
{}_i \langle k| \hat{D}_i (\beta_i/\sqrt{N}) \hat{N}_{BS} |\Psi\rangle_0 
\right|^2
\nonumber\\ 
& \le & \frac{ 
\langle \Psi_{\beta_i} | \hat{a}_0^{\dagger k} \hat{a}_0^k 
|\Psi_{\beta_i} \rangle 
}{N^k k!} 
+ O \left( \frac{1}{N^{k+1}} \right) 
\nonumber\\ & \le & 
C_k^{\rm max} / N^k + O ( 1/N^{k+1} ), 
\end{eqnarray}
where $|\Psi_{\beta_i} \rangle \equiv \hat{D}(C_{\beta_i}) |\Psi\rangle$, 
whose probability distribution still decreases exponentially 
in number basis (see Appendix A), and 
$C_k^{\rm max}$ is the maximum value of 
$\langle \Psi_{\beta_i} | \hat{a}_0^{\dagger k} 
\hat{a}_0^k |\Psi_{\beta_i} \rangle / k!$
for all $i$ and possible inputs \cite{C_k^max}. 
The probability of resulting the failure event $P_{fail}$ is then 
bounded as 
\begin{eqnarray}
\label{eq:P_failure}
P_{fail} 
& \le &
( C_2^{\rm max} / N^2 + O (1/N^3) ) \times N 
\nonumber\\ & = & 
C_2^{\rm max} / N + O (1/N^2), 
\end{eqnarray}
which implies that $P_{fail}$ approaches to zero 
in the limit of large $N$, at least with the order of $1/N$.

Even if the detection is successful, 
the conditional states get slightly non-orthogonal 
after each measurement step. 
To see this, we revisit 
the first beamsplitter $\hat{B}_{1,0}(\theta_1)$. 
Let us describe the states after beamsplitting 
such that the orthogonal and non-orthogonal parts are separated as 
\begin{eqnarray}
\label{eq:B|0>|psi>}
\hat{B}_{1,0} (\theta_1) |0\rangle |\Psi\rangle & = & 
|0\rangle |\eta_0\rangle + N^{-1/2} |1\rangle |\eta'_1\rangle
+ N^{-1} |2\rangle |\eta_2\rangle + \cdots 
\nonumber\\ & = & 
|0\rangle |\eta_0\rangle 
+ N^{-1/2} |1\rangle |\eta_1\rangle 
+ N^{-3/2} |1\rangle |\eta_r\rangle 
\nonumber\\ & & 
+ \sum_{k=2}^\infty N^{-k/2} 
|k\rangle|\eta_k\rangle, \\
\label{eq:B|0>|phi>}
\hat{B}_{1,0} (\theta_1) |0\rangle |\Phi\rangle 
& = & 
|0\rangle |\nu_0\rangle 
+ N^{-1/2} |1\rangle |\nu_1\rangle 
+ N^{-3/2} |1\rangle |\nu_r\rangle 
\nonumber\\ & & 
+ \sum_{k=2}^\infty N^{-k/2}
|k\rangle|\nu_k\rangle ,
\end{eqnarray}
where the first two terms exactly satisfy the orthogonality 
$\langle\eta_0|\nu_0\rangle + \langle\eta_1|\nu_1\rangle /N = 0$ 
and the last terms represent the multi-photon reflection terms. 
Here, $|\eta_0\rangle = \sum^{\infty}_{m=0} c_m (1-1/N)^{m/2} |m\rangle$, 
$N^{-1/2} |\eta'_1\rangle 
= \sum^{\infty}_{m=1} c_m (m/N)^{1/2} (1-1/N)^{(m-1)/2} 
|m-1\rangle$, 
$N^{-1/2} |\eta_1\rangle 
= \sum^{\infty}_{m=1} c_m (1-(1-1/N)^m)^{1/2} |m-1\rangle$, 
and $N^{-3/2} |\eta_r\rangle 
= N^{-1/2}(|\eta'_1\rangle - |\eta_1\rangle)$ 
($|\nu_n\rangle$'s are also obtained by replacing $c_m$ with $d_m$). 
The terms $|\eta_r\rangle$, $|\nu_r\rangle$ and that 
for multi-photon reflections, 
which have been neglected in the previous discussion, 
cause the residual non-orthogonality. Note that 
the leading terms of all vectors $|\eta_k\rangle$'s and $|\nu_k\rangle$'s 
are independent of $N$. 
Denote the $i$-th measurement operation as
\begin{equation}
\label{eq:Kraus_op}
\frac{
{}_i \langle k| 
\hat{D}_i(\beta_i/\sqrt{N}) \hat{B}_{i,0}(\theta_i) |0\rangle_i 
|\Psi\rangle }{|
{}_i \langle k| \hat{D}_i(\beta_i/\sqrt{N}) \hat{B}_{i,0}(\theta_i) |0\rangle_i
|\Psi\rangle |
}
\equiv \hat{E}^{(i)}_k |\Psi\rangle. 
\end{equation}
Then the conditional outputs after 
detecting zero and one photons at the first measurement 
are given by 
\begin{eqnarray}
\label{eq:output_0}
\hat{E}_0^{(1)} |\Psi\rangle
& = & 
\mathcal{N}_0 
\left\{ |\eta_0\rangle - \frac{\beta_1^*}{N} |\eta_1\rangle 
+ \frac{1}{N^2} |\eta_{R_0}^{(1)} \rangle
\right\} ,
\\
\label{eq:output_1}
\hat{E}_1^{(1)} |\Psi\rangle
& = & 
\mathcal{N}_1
\left\{ \beta_1 |\eta_0\rangle + |\eta_1\rangle 
+ \frac{1}{N} |\eta_{R_1}^{(1)} \rangle 
\right\} , 
\end{eqnarray}
respectively, where $\mathcal{N}_0$ and $\mathcal{N}_1$ 
are the normalization factors 
and the third terms $|\eta_{R_i}^{(1)}\rangle$'s ($i=0,1$) 
come from $|\eta_r\rangle$ and $|\eta_k\rangle$'s for $k\ge2$, 
and the terms in Eqs.~(\ref{eq:approx_displaced_pc_0}) 
and (\ref{eq:approx_displaced_pc_1}) 
whose order is higher than $1/N^{1/2}$. 
The same outputs are obtained for $|\Phi\rangle$ 
by replacing $|\eta_n\rangle$ with $|\nu_n\rangle$. 
The first two terms in Eqs.~(\ref{eq:output_0}) and (\ref{eq:output_1})
can be exactly orthogonal to those of $|\Phi\rangle$ 
by choosing $\beta_1/\sqrt{N} = (1-\sqrt{1+X^2})e^{i\omega}/X$, 
where X is obtained by substituting $|\eta_0\rangle$, $|\eta_1\rangle$, 
$|\nu_0\rangle$ and $|\nu_1\rangle$, 
appearing in Eqs.~(\ref{eq:B|0>|psi>}) and (\ref{eq:B|0>|phi>}), 
into Eq.~(\ref{eq:X}). 
Since $X \propto 1/\sqrt{N}$ as mentioned above, 
this choice of $\beta_1$ always satisfy the constraint 
on the average power of the local oscillator, 
$|\beta_1|^2 \le |C_{\beta_1}|^2 + O(1/N)$. 
However, we have to care of the fact that, 
in both events, the {\it total} conditional states 
in Eqs.~(\ref{eq:output_0}) and (\ref{eq:output_1}) 
are no longer orthogonal  due to their third terms.

Now, suppose that the same strategy is applied to the choice of $\beta_2$ 
for the second measurement step. 
After the second measurement, the states are mapped into 
the new one with orthogonal and non-orthogonal terms, where the latter 
has two parts, i.e. contributions from the first and second measurements. 
Note that the leading order of prefactors of $|\eta_{R_k}^{(1)}\rangle$ 
with respect to $1/N$ does not change during the measurement process, as also the leading factors of  $|\Psi\rangle$ does not change in the mapping
in Eqs.~(\ref{eq:output_0}) and (\ref{eq:output_1}).  
Eventually, after repeating $N-1$ measurement steps in a similar way, 
if all the photon counters detected zero or one photons, 
one obtains the conditional output consists of the orthogonal term and 
$N-1$ non-orthogonal terms stemmed from each measurement as 
\begin{eqnarray}
\label{eq:N-1th_meas}
|\Psi^{(N-1)}\rangle & = & 
\hat{E}^{(N-1)} \cdots \hat{E}^{(1)} |\Psi\rangle
\nonumber\\
& = & 
|\eta^{(N-1)}\rangle
+ \frac{1}{N^2} \sum_{x=1}^{I^{(N-1)}} |H_0^{(i_x)}\rangle
+ \frac{1}{N} \sum_{y=1}^{J^{(N-1)}} |H_1^{(j_y)}\rangle, 
\nonumber\\
\end{eqnarray}
where the first term is 
exactly orthogonal to that of $|\Phi^{(N-1)}\rangle$, while 
$|H_k^{(l)}\rangle$ is the residual non-orthogonal term 
coming from $|\eta_{R_k}^{(l)}\rangle$. 
$I^{(N-1)}$ and $J^{(N-1)}$ are the numbers of the events of 
detecting zero and one photon, respectively, and thus 
$I^{(N-1)} + J^{(N-1)} = N-1$.

Let us denote the final $N$-th measurement by $|D_k\rangle \equiv 
\hat{D}^{\dagger} (\beta_N/\sqrt{N}) |k\rangle$ ($k=0,1$). 
Suppose that $\beta_N$ is designed such that 
$|D_0\rangle$ and $|D_1\rangle$ are the same as the orthogonal terms in 
$|\Psi^{(N-1)}\rangle$ and $|\Phi^{(N-1)}\rangle$, respectively, 
up to the order of $1/N^{1/2}$ (the higher order terms 
contribute to the detection error). 
Then the error probability 
$P_{err}^{D_1} = |\langle D_1|\Psi^{(N-1)}\rangle|^2$ 
is given by 
\begin{eqnarray} 
\label{eq:error_prob1} 
P_{err}^{D_1} & = & 
\left| 
\sum_{x=1}^{I^{(N)}} \frac{\langle D_1|H_0^{(i_x)}\rangle}{N^2} 
+ \sum_{y=1}^{J^{(N)}} \frac{\langle D_1|H_1^{(j_y)}\rangle}{N} 
\right|^2 . 
\end{eqnarray} 
where $I^{(N)} + J^{(N)} = N$. 
The leading order of $\langle D_1|H_k^{(j)}\rangle$ is independent 
of $N$ for every $j$ and $k$.

\begin{figure}
\begin{center}
\includegraphics[width=1\linewidth]{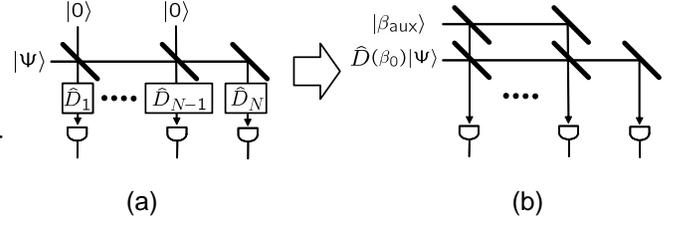}   %
\caption{\label{fig:total_power}
The original scheme (a) can be transformed into (b) 
where the total input photon number is the sum of 
those of two input states. 
}
\end{center}
\end{figure}

One can estimate the order of $J^{(N)}$ by counting 
the total amount of photons put into the system since 
the number of the total photon is equal to that of detectors. 
Although photons are supplied by the input state 
and $N$ displacements in the original configuration, 
one can simplify it into the one 
with only two inputs, $\hat{D}(\beta_0) |\Psi\rangle$ and 
the coherent state $|\beta_{\rm aux}\rangle$, by adding some linear optics  
as illustrated in Fig.~\ref{fig:total_power}. 
Here, with the relation 
$\hat{D}_A (\alpha) \hat{D}_B (\beta) \hat{B}_{AB}(\theta) 
= \hat{B}_{AB}(\theta) \hat{D}_A (\alpha\cos\theta - \beta\sin\theta) 
\hat{D}_B (\alpha\sin\theta + \beta\cos\theta)$, one finds 
$|\beta_0|^2 = |\sum_{i=1}^N \beta_i /N|^2$ and 
$|\beta_{\rm aux}|^2 = \sum_{i=1}^N |\beta_i|^2/N - |\beta_0|^2$, 
where these are bounded as $|\beta_0|^2 = C_0 + O(1/N)$ and 
$|\beta_{\rm aux}|^2 = C_{\rm aux} + O(1/N)$ due to 
the constraint on $|\beta_i|^2$'s. 
$C_0$ and $C_{\rm aux}$ are constants independent of $N$.

The probability of having $n$ photons in total is given by 
$P(n) = \sum_{m=0}^n P_{\rm sig} (n-m) P_{\rm aux} (m)
= C_P e^{-n x} + O(1/N)$. 
Here the photon number statistics of two inputs, 
$P_{\rm sig}(m)$ and $P_{\rm aux}(m)$ are exponential and Poissonian, 
which easily implies that $P(n)$ decreases exponentially 
with resepect to $n$ (see Appendix C). 
Therefore, one can bound $J^{(N)}$ by some constant $C_J$ 
with exponentially small exception as 
\begin{eqnarray} 
\label{eq:J} 
&&{\rm Prob} \left[ J^{(N)} \le C_J + O(1/N) + N\epsilon \right] 
\nonumber\\&&
\ge 1 - C_P \exp[ -(C_J+O(1/N)+N\epsilon)] + O(1/N) 
\nonumber\\&& 
= 1-\tilde{C}_P e^{-N\epsilon} + O(1/N)
\end{eqnarray}
where $\epsilon$ can be arbitrarily small for large $N$. 
Eventually, substituting it and $I^{(N)} \le N$ into 
Eq.~(\ref{eq:error_prob1}), one obtains 
\begin{eqnarray} 
\label{eq:error_prob1-0} 
P_{err}^{D_1} & = & 
\left| 
\frac{I^{(N)}}{N^2} \langle D_1|H_0\rangle_{\rm av} 
+ \frac{J^{(N)}}{N} \langle D_1|H_1\rangle_{\rm av} \right|^2
\nonumber\\ & \le &
C_E/N^2 + O(1/N^3) + \epsilon O(1/N) + \epsilon^2,
\end{eqnarray} 
where $\langle D_1|H_k\rangle_{\rm av} 
= \sum_i \langle D_1|H_k^{(i)}\rangle / L$ 
($L=I^{(N)}$ and $J^{(N)}$ for $k=0,\,1$, respectively), 
and $C_E$ is some constant independent of $N$. 
In a similar manner, the same bound is derived for 
$P_{err}^{D_0} = |\langle D_0|\Phi^{(N-1)}\rangle|^2$. 
Then, summing over all detection patterns, 
the average error probability 
is bounded as 
\begin{eqnarray}
\label{eq:error_prob_av}
P_{err}^{\rm tot} & = & 
\sum^{\rm success} P(\sharp) P^{succ}_{err}(\sharp) 
+ \sum^{\rm failure} P(\sharp) P_{fail}(\sharp) 
\nonumber\\
& \le & 
\left( 1 - \frac{C_2^{\rm max}}{N} \right) 
\frac{P_{err}^{D_0} + P_{err}^{D_1}}{2}
+ \frac{C_2^{\rm max}}{N} + O \left( \frac{1}{N^2} \right)
\nonumber\\ & \le & 
C/N + O ( 1/N^2 ) 
+ O(1/N) \epsilon + \epsilon^2,
\end{eqnarray}
where $C$ is some constant and $P(\sharp)$ is the probability to 
observe the measurement sequence pattern $\sharp$. 
As a consequence, 
in the limit of $N\to\infty$, one can discriminate 
$|\Psi\rangle$ and $|\Phi\rangle$ with unit probability.

Finally, we discuss the case $|\langle\eta_0|\nu_1\rangle|^2 
- |\langle\eta_1|\nu_0\rangle|^2 = 0$ in Eq.~(\ref{eq:X}), 
in which the desirable local measurement can not be implemented 
by a displacement and photon counting. 
Here, let us consider the projection measurement 
consisting of slightly perturbed vectors 
$|\Pi_0\rangle = \sqrt{1-\delta}|\Psi\rangle - \sqrt{\delta}|\Phi\rangle$
and 
$|\Pi_1\rangle = \sqrt{1-\delta}|\Phi\rangle + \sqrt{\delta}|\Psi\rangle$ 
with a perturbation parameter $\delta$. 
One can design such a measurement by the previous strategy 
with the total error probability of 
$P_{err}^{\rm tot} =  
C/N^{1-2\Delta} + O (1/N^{2-3\Delta}) 
+ O (1/N^{1-3\Delta/2}) \epsilon + O (N^\Delta) \epsilon^2$,
where $\Delta = - \log_N \delta$. 
This device can discriminate 
the original states 
$|\Psi\rangle$ and $|\Phi\rangle$ 
with the average error probability of 
\begin{eqnarray}
\label{eq:error_prob_av_special2}
P_{err}^{\rm av} & = & 1 - 
(1-P_{err}^{\rm tot})
(|\langle\Pi_0|\Psi\rangle|^2 + |\langle\Pi_1|\Phi\rangle|^2)/2  
\nonumber\\ & = & 
C_1/N^{\Delta} + O (1/N^{2\Delta}) 
+ C_2/N^{1-2\Delta} + O (1/N^{2-3\Delta}) 
\nonumber\\ & & 
+ O (1/N^{1-3\Delta/2}) \epsilon + O (N^\Delta) \epsilon^2
\end{eqnarray}
In the asymptotic limit of large $N$, this is minimized with $\Delta = 1/3$ 
and then we obtain 
$P_{err}^{\rm av} = C/N^{1/3} + O(1/N^{2/3}) 
+ O(1/N^{1/2}) \epsilon + O(N^{1/3}) \epsilon^2$
which still converges to zero.

In summary, we have proved that arbitrary two orthogonal pure states 
can be perfectly discriminated by linear optics tools without using 
any non-classical ancillary states in the asymptotic limit of 
$N\to\infty$ where $N$ is the number of the detections and feedforwards. 
It implies that, in principle, one can implement arbitrary projection 
measurement in any two-dimensional signal space by these tools. 
The resources discussed here are mostly available with current technology. 
We also showed a concrete designing strategy of a linear optics circuit 
to attain this bound for a given $N$ and thus 
it can be directly applied for various quantum information protocols 
that require binary projection measurements. 
The remaining question is whether one can apply a same approach to 
the problem of more than three states discrimination.

We thank M.~Ban, D.~Berry, K.~Tamaki, and P.~van Loock 
for valuable discussions and comments. 
M.T. also acknowledges a kind hospitality at the QIT group 
in Universit\"{a}t Erlangen-N\"{u}rnberg. 
This work was supported by the DFG under the Emmy-Noether
program, the EU FET network RAMBOQ and the network of competence QIP of
the state of Bavaria.

\appendix
\section{Photon number statistics of the displaced state}
In this appendix, we show that if the photon number distribution of 
the initial state is exponential, then that of its displaced state is 
also bounded by exponentially decreasing function. 
For this purpose we use the following three formulae;

{\bf (1) The number basis components of the displacement operator 
\cite{Cahill69};}
\begin{eqnarray}
\label{eq:A_1}
\langle n| \hat{D}(\xi) |m\rangle & = & 
\sqrt{\frac{m!}{n!}} \xi^{n-m} e^{-|\xi|^2/2} L_m^{(n-m)} (|\xi|^2), 
\end{eqnarray}
for $(n \ge m)$ and
\begin{eqnarray}
\langle n| \hat{D}(\xi) |m\rangle & = & 
\sqrt{\frac{n!}{m!}} (-\xi^*)^{m-n} e^{-|\xi|^2/2} L_n^{(m-n)} (|\xi|^2), 
\nonumber\\
\end{eqnarray}
for $(n \le m)$, where $L_n^{(l)} (x)$ is the associated Laguerre polynomial  
defined by 
\begin{equation}
\label{eq:A_1_1}
L_n^{(l)} (x) = \sum_{k=0}^n {n+l \choose n-k} 
\frac{(-x)^k}{k!} ,
\end{equation}
where $L_n^{(0)}(x) = L_n(x)$ is the Laguerre polynomial and 
\begin{equation}
\label{eq:A_1_2}
\frac{{\rm d}^l}{{\rm d} x^l} L_n (x) = (-1)^l L_{n-l}^{(l)} (x). 
\end{equation}

{\it Proof}. We basically follow the proof given in \cite{Barnett97}. 
To calculate $\langle n| \hat{D}(\xi) |m\rangle$, 
it is helpful to see $\langle n| \hat{D}(\xi) |n\rangle$, 
which is given by 
\begin{eqnarray}
\label{eq:A_1_3}
\langle n| \hat{D}(\xi) |n\rangle & = & 
\langle n| \exp \left( \xi\hat{a}^\dagger - \xi^*\hat{a} \right) 
|n\rangle 
\nonumber\\ & = & 
e^{-|\xi|^2/2} \langle n| e^{\xi\hat{a}^\dagger} 
e^{- \xi^*\hat{a}} |n\rangle 
\nonumber\\ & = & 
\sum_{l=0}^\infty \sum_{m=0}^\infty e^{-|\xi|^2/2} 
\frac{\xi^l (-\xi^*)^m}{l!m!} 
\langle n| \hat{a}^{\dagger \, l} \hat{a}^m |n\rangle 
\nonumber\\ & = & 
\sum_{l=0}^\infty \sum_{m=0}^\infty e^{-|\xi|^2/2} 
\frac{\xi^l (-\xi^*)^m}{l!m!} 
\nonumber\\ && 
\langle n-l| \sqrt{\frac{n!}{(n-l)!}} 
\sqrt{\frac{n!}{(n-m)!}} |n-m\rangle 
\nonumber\\ & = & 
e^{-|\xi|^2/2} \sum_{m=0}^n {n \choose m} 
\frac{(-|\xi|^2)^m}{m!} 
\nonumber\\ & = & 
e^{-|\xi|^2/2} L_n (|\xi|^2). 
\end{eqnarray}
Then we obtain 
\begin{eqnarray}
\label{eq:A_1_4}
& & \langle n| \hat{D}(\xi) |n-l\rangle 
\nonumber\\ && =  
e^{-|\xi|^2/2} \langle n| e^{\xi\hat{a}^\dagger} 
e^{- \xi^*\hat{a}} \hat{a}^l |n\rangle 
\sqrt{\frac{(n-l)!}{n!}} 
\nonumber\\ && =  
e^{-|\xi|^2/2} \sqrt{\frac{(n-l)!}{n!}} 
\left( -\frac{\partial}{\partial \xi^*} \right)^l 
\langle n| e^{\xi\hat{a}^\dagger} e^{- \xi^*\hat{a}} |n\rangle 
\nonumber\\ && =  
e^{-|\xi|^2/2} \sqrt{\frac{(n-l)!}{n!}} 
(-\xi)^l \left( \frac{\partial}{\partial |\xi|^2} \right)^l 
L_n (|\xi|^2) .
\end{eqnarray}
Therefore, replacing $n-l$ with $m$ in Eq.~(\ref{eq:A_1_4}) 
with Eq.~(\ref{eq:A_1_2}), 
we can derive Eq.~(\ref{eq:A_1}).

{\bf (2) Bound on the associated Laguerre polynomials 
\cite{HandbookMath};}
\begin{eqnarray}
\label{eq:A_2}
\left| L_n^{(a)} (x) \right| \le 
{a+n \choose n} e^{x/2}, 
\end{eqnarray}
where $x \ge 0$ and $a$ is an integer. 

{\it Proof}. From Eq.~(\ref{eq:A_1_3}), the absolute value of 
the Laguerre polynomial $L_n (x)$ with $x\ge0$ is bounded by 
\begin{eqnarray}
\label{eq:A_2_1}
|L_n (x)| & = & e^{x/2} 
\left| \langle n|\hat{D}(x^{1/2}) |n\rangle \right| 
\nonumber\\ & \le & 
e^{x/2} .
\end{eqnarray}

To extend it to the associated Laguerre polynomial, 
we use the relation 
\begin{equation}
\label{eq:A_2_2}
L_n^{(a)}(x) = \sum_{k=0}^n {a+k-1 \choose a-1} L_{n-k}(x) , 
\end{equation}
which can be derived as 
\begin{eqnarray}
\label{eq:A_2_3}
& = & 
\sum_{k=0}^n {a+k-1 \choose a-1} 
\sum_{l=0}^{n-k} {n-k \choose l} \frac{(-x)^l}{l!} 
\nonumber\\ & = & 
\sum_{l=0}^n \sum_{k=0}^{n-l} 
{a+k-1 \choose a-1} {n-k \choose l} \frac{(-x)^l}{l!} 
\nonumber\\ & = & 
\sum_{l=0}^n {n+a \choose n-l} \frac{(-x)^l}{l!} 
\nonumber\\ & = & 
L_n^{(a)} (x) ,
\end{eqnarray}
where the formula 
\begin{equation}
\label{eq:A_2_4}
\sum_{k=0}^{n-l} {a+k \choose a} {n-k \choose l} 
= {n+a+1 \choose n-l}  
\end{equation}
has been utilized. 
Eventually, Eqs.~(\ref{eq:A_2_1}), (\ref{eq:A_2_2}) 
and (\ref{eq:A_2_4}) imply 
\begin{eqnarray}
\label{eq:A_2_5}
\left| L_n^{(a)}(x) \right| & \le & 
\sum_{k=0}^n {a+k-1 \choose a-1} \left| L_{n-k}(x) \right|
\nonumber\\ & \le & 
\sum_{k=0}^n {a+k-1 \choose a-1} e^{x/2} 
= {a+n \choose a} e^{x/2} . \nonumber\\ 
\end{eqnarray}

{\bf (3) Inequality for the binomial distribution
\cite{GenIneq};}
\begin{eqnarray}
\label{eq:A_5}
{n \choose \nu} y^\nu (1-y)^{n-\nu} \le 
\exp \left[ -2n (y-\nu/n)^2 \right], 
\end{eqnarray}
where $n > \nu$ and $0<y<1$. 

{\it Proof}. Define $q=\nu/n$ and 
\begin{eqnarray}
\label{eq:A_5_1}
f(y) & = & y^\nu (1-y)^{n-\nu} e^{-2n(x-q)^2}, 
\nonumber\\ 
F(y) & = & n^{-1} \ln f(x). 
\end{eqnarray}
Then 
\begin{eqnarray}
\label{eq:A_5_2}
F(y) & = & q \ln y + (1-q) \ln (1-x) + 2(x-q)^2, 
\nonumber\\ 
F'(y) & = & \frac{(q-y)(1-2y)^2}{y(1-y)}, 
\end{eqnarray}
and thus $F(y)$ takes its maximum at $y=q$. 
Also, the same for $f(y)$. 
Therefore, $f(y) \le f(q)$, i.e. 
\begin{equation}
\label{eq:A_5_3}
y^\nu (1-y)^{n-\nu} e^{2n(y-q)^2} \le q^\nu (1-q)^{n-\nu}, 
\end{equation}
and thus 
\begin{equation}
\label{eq:A_5_4}
{n \choose \nu} y^\nu (1-y)^{n-\nu} e^{2n(y-q)^2} 
\le {n \choose \nu} q^\nu (1-q)^{n-\nu} \le 1, 
\end{equation}
which completes the proof.

\begin{widetext}
{\bf Derivation of the displaced state.}
Now we derive the main statement of this appendix. 
We assume that $|\Psi\rangle$ can be written as 
\begin{equation}
|\Psi\rangle = \sum_{m=0}^\infty \tilde{c}_m e^{-mx/2} |m\rangle. 
\end{equation}
Now, let us calculate $\langle n| \hat{D}(\beta) |\Psi\rangle$. 
\begin{eqnarray}
\label{eq:A_6}
&& \left| \langle n| \hat{D}(\beta) |\Psi\rangle \right| = 
\left| \sum_{m=0}^\infty \tilde{c}_m e^{-mx/2} 
\langle n| \hat{D}(\beta) |m \rangle \right|
\nonumber\\ && \le  
\sum_{m=0}^\infty \left| \tilde{c}_m e^{-mx/2} 
\langle n| \hat{D}(\beta) |m \rangle \right|
\nonumber\\ && \le  
\sum_{m=0}^n |\tilde{c}_m| e^{-mx/2} \left( \frac{m!}{n!} \right)^{1/2} 
|\beta|^{n-m} e^{-|\beta|^2/2} 
\left| L_m^{(n-m)} (|\beta|^2) \right| 
+ \sum_{m=n+1}^\infty |\tilde{c}_m| e^{-mx/2} 
\left( \frac{n!}{m!} \right)^{1/2} 
|\beta|^{m-n} e^{-|\beta|^2/2} 
\left| L_n^{(m-n)} (|\beta|^2) \right| 
\nonumber\\ && \le 
|\tilde{c}_{\rm max}| \sum_{m=0}^n e^{-mx/2} 
\left( \frac{m!}{n!} \right)^{1/2} |\beta|^{n-m} {n \choose m}
+ |\tilde{c}_{\rm max}| \sum_{m=n+1}^\infty e^{-mx/2} 
\left( \frac{n!}{m!} \right)^{1/2} |\beta|^{m-n} {m \choose n}
\nonumber\\ && =  
|\tilde{c}_{\rm max}| \sum_{m=0}^n e^{-nx/2} 
\left\{ {n \choose m} \frac{(|\beta|^2 e^x)^{n-m}}{(n-m)!} \right\}^{1/2} 
+ |\tilde{c}_{\rm max}| \sum_{m=n+1}^\infty e^{-nx/2} 
\left\{ {m \choose n} \frac{(|\beta|^2 e^{-x})^{m-n}}{(m-n)!} \right\}^{1/2} 
\nonumber\\ && \le  
\left\{ n e^{-nx} |\tilde{c}_{\rm max}|^2 \sum_{m=0}^n {n \choose m} 
\frac{(|\beta|^2 e^x)^{n-m}}{(n-m)!} \right\}^{1/2} 
+ \left\{ n e^{-nx} |\tilde{c}_{\rm max}|^2 \sum_{m=n+1}^\infty {m \choose n} 
\frac{(|\beta|^2 e^{-x})^{m-n}}{(m-n)!} \right\}^{1/2} .
\end{eqnarray}
The last line follows from the Cauchy-Schwarz inequality. 

Introducing a real parameter $q$ which satisfies $e^{-x}<q<1$, 
the first term of Eq.~(\ref{eq:A_6}) is then bounded as
\begin{eqnarray}
\label{eq:A_7}
&& \sqrt{n} e^{-nx/2} |\tilde{c}_{\rm max}| \left\{ \sum_{m=0}^n {n \choose m} 
\frac{(|\beta|^2 e^x)^{n-m}}{(n-m)!} \right\}^{1/2}
\nonumber\\ && = 
\sqrt{n} e^{-nx/2} |\tilde{c}_{\rm max}| \left\{ \sum_{m=0}^n {n \choose m} 
q^m (1-q)^{n-m} 
\frac{1}{(n-m)!} \left(\frac{q|\beta|^2 e^x}{1-q}\right)^{n-m} 
q^{-n} \right\}^{1/2} 
\nonumber\\ && \le 
\sqrt{n} \left( \frac{e^{-x}}{q} \right)^{n/2} |\tilde{c}_{\rm max}| 
\left\{ \sum_{m=0}^n {n \choose m} q^m (1-q)^{n-m} 
\exp\left[ \frac{q|\beta|^2 e^x}{1-q} \right] \right\}^{1/2}
\nonumber\\ &&
= \sqrt{n} \left( \frac{e^{-x}}{q} \right)^{n/2} |\tilde{c}_{\rm max}| 
\exp\left[ \frac{q|\beta|^2 e^x}{2(1-q)} \right], 
\nonumber\\ 
\end{eqnarray}
and thus it decreases exponentially as $n$ increases. 
Also, for the second term, one obtains 
\begin{eqnarray}
\label{eq:A_8}
&& \sqrt{n} e^{-nx/2} |\tilde{c}_{\rm max}| \left\{ \sum_{m=n+1}^\infty 
{m \choose n} \frac{(|\beta|^2 e^{-x})^{m-n}}{(m-n)!} \right\}^{1/2}
\nonumber\\ && = 
\sqrt{n} e^{-nx/2} |\tilde{c}_{\rm max}| \left\{ \sum_{m=n+1}^\infty 
{m \choose n} q^n (1-q)^{m-n} 
\frac{1}{(m-n)!} \left(\frac{q|\beta|^2 e^{-x}}{1-q}\right)^{m-n} 
q^{-n} \right\}^{1/2} 
\nonumber\\ && \le 
\sqrt{n} \left( \frac{e^{-x}}{q} \right)^{n/2} |\tilde{c}_{\rm max}| 
\left\{ \sum_{m=n+1}^\infty {m \choose n} q^n (1-q)^{m-n} 
\exp\left[ \frac{q|\beta|^2 e^{-x}}{1-q} \right] \right\}^{1/2} 
\nonumber\\ && \le 
\sqrt{n} \left( \frac{e^{-x}}{q} \right)^{n/2} |\tilde{c}_{\rm max}| 
\exp\left[ \frac{q|\beta|^2 e^{-x}}{2(1-q)} \right] 
\left\{ 
\sum_{m=n+1}^\infty \exp\left[ -2m \left(q-\frac{n}{m}\right)^2 \right]
\right\}^{1/2} . 
\nonumber\\ 
\end{eqnarray}
Since the last exponential term decreases exponentially as $m$ increase 
at least in the limit of $m \gg n$, the sum always converges 
within a finite value, which means that  
Eq.~(\ref{eq:A_8}) itself also decreases exponentially as $n$ increases. 
As a consequence, these results imply that 
Eq.~(\ref{eq:A_6}) decreases exponentially as $n$ increases.

\section{Derivation of the inequality (\ref{eq:P_k_average})}
In this appendix, we derive the inequality (\ref{eq:P_k_average}). 
\begin{eqnarray}
\label{eq:P_k_average_appendix}
P_k^{(i)} & = & \left| 
{}_i \langle k| \hat{D}_i (\beta_i/\sqrt{N}) \hat{N}_{BS} |\Psi\rangle_0 
\right|^2
\nonumber\\ & = & 
\left|
{}_i \langle k| \hat{D}_i (\beta_i/\sqrt{N}) \hat{B}_{N-1,0} (\theta_{N-1}) 
\cdots \hat{B}_{i,0} (\theta_i) \cdots \hat{B}_{1,0} (\theta_1) 
|0\rangle^{\otimes N-1} |\Psi\rangle_0 
\right|^2
\nonumber\\ & = & \left|
{}_i \langle k| \hat{D}_i (\beta_i/\sqrt{N}) \hat{B}_{N-1,0} (\theta_{N-1}) 
\cdots \hat{B}_{i+1,0} (\theta_{i+1}) \hat{B}_{i-1,0} (\theta_i) 
\cdots \hat{B}_{1,0} (\theta_2) \hat{B}_{i,0} (\theta_1) 
|0\rangle^{\otimes N-1} |\Psi\rangle_0 
\right|^2
\nonumber\\ & = & \left|
{}_i \langle k| \hat{D}_i (\beta_i/\sqrt{N}) \hat{B}_{i,0} (\theta_1) 
|0\rangle_i |\Psi\rangle_0 
\right|^2
\nonumber\\ & = & \left|
{}_i \langle k| e^{-|\beta_i|^2/2N} 
e^{\beta_i \hat{a}_i^\dagger/\sqrt{N}} e^{-\beta_i^* \hat{a}_i/\sqrt{N}} 
e^{- \hat{a}_i^\dagger \hat{a}_0 \tan\theta_1}  
e^{- \ln \cos\theta_1 
(\hat{a}_i^\dagger \hat{a}_i-\hat{a}_0^\dagger \hat{a}_0) } 
e^{\hat{a}_i \hat{a}_0^\dagger \tan\theta_1}  
|0\rangle_i |\Psi\rangle_0 \right|^2
\nonumber\\ & = & \left| 
{}_i \langle k| e^{-|\beta_i|^2/2N} 
e^{\beta_i \hat{a}_i^\dagger/\sqrt{N}} 
e^{-\hat{a}_i^\dagger \hat{a}_0 / \sqrt{N-1}} 
e^{\beta_i^* \hat{a}_0 / \sqrt{N(N-1)}} 
e^{-\beta_i^* \hat{a}_i / \sqrt{N}} 
e^{\ln \sqrt{1-1/N} \hat{a}_0^\dagger \hat{a}_0} 
|0\rangle_i |\Psi\rangle_0 \right|^2
\nonumber\\ & = & \left|
e^{-|\beta_i|^2/2N} \left\{
\sum_{j=0}^k {}_i \langle k-j| \frac{1}{j!} \sqrt{\frac{k!}{(k-j)!}}
\left( \frac{\beta_i}{\sqrt{N}} - \frac{\hat{a}_0}{\sqrt{N-1}} \right)^j 
\right\}
e^{\beta_i^* \hat{a}_0 / \sqrt{N(N-1)}} 
e^{\ln \sqrt{1-1/N} \hat{a}_0^\dagger \hat{a}_0} 
|0\rangle_i |\Psi\rangle_0 \right|^2
\nonumber\\ & = & \left|
\frac{e^{-|\beta_i|^2/2N}}{\sqrt{k!}} 
e^{\hat{a}_0^{\dagger}\hat{a}_0 \ln \sqrt{1-1/N}} 
\left( \frac{\beta_i - \hat{a}_0}{\sqrt{N}} \right)^k
e^{\beta_i^* \hat{a}_0 /N} 
|\Psi\rangle 
\right|^2 
\nonumber\\ 
& \le & \frac{ 
\langle \Psi_{\beta} | \hat{a}_0^{\dagger k} \hat{a}_0^k 
|\Psi_{\beta} \rangle 
}{N^k k!} 
+ O \left( \frac{1}{N^{k+1}} \right) 
\end{eqnarray}
\end{widetext}
where $\cos\theta_1 = \sqrt{1-1/N}$, $\sin\theta_1 = 1/\sqrt{N}$, 
$|\beta_i|^2 \le |C_{\beta_i}|^2 + O(1/N)$ and 
$|\Psi_{\beta_i} \rangle = \hat{D}(C_{\beta_i}) |\Psi\rangle$. 
We have used the relation 
$e^{\alpha \hat{a}_j} e^{\beta \hat{a}_j^\dagger} 
= e^{\beta \hat{a}_j^\dagger} e^{\alpha \hat{a}_j} e^{\alpha\beta}$ 
from line 5 to 6, and 
$e^{\phi \hat{a}_j^\dagger \hat{a}_j} \hat{a}_j 
e^{-\phi \hat{a}_j^\dagger \hat{a}_j} = \hat{a}_j e^{-\phi}$ 
from line 7 to 8, 
where $\alpha$, $\beta$, and $\phi$ are complex numbers. 
These relations are directly obtained from the commutation relation 
$[\hat{a}_j, \hat{a}_j^\dagger] = 1$. 

The remaining task is to show that 
$\langle\Psi_{\beta_i}| \hat{a}^{\dagger \, k} \hat{a}^k 
|\Psi_{\beta_i}\rangle/k!$ 
is always finite, i.e.  
\begin{eqnarray}
\label{eq:B_1}
\frac{\langle\Psi_{\beta_i}| \hat{a}^{\dagger \, k} \hat{a}^k 
|\Psi_{\beta_i}\rangle
}{N^k k!} 
\le \frac{C_k^{\rm max}}{N^k} ,
\end{eqnarray}
with a constant $C_k^{\rm max}$. 
Here, we replace $\beta_i$ by $\beta$ for simplicity. 
As shown in Appendix A, the photon number distribution of 
$|\Psi_\beta\rangle$ 
decreases exponentially. Denote
\begin{eqnarray}
\label{eq:B_2}
|\Psi_\beta\rangle 
\equiv \sum_{m=0}^\infty \tilde{b}_m e^{-mx/2} |m\rangle .
\end{eqnarray}
The absolute of complex coefficients $\tilde{b}_m$'s are always 
in between 0 and some constant due to the normalization constraint 
and let us denote the constant as $|\tilde{b_{\rm max}}|$. 
Then one has 
\begin{eqnarray}
\label{eq:B_3}
\hat{a}^k |\Psi_\beta\rangle & = & 
\sum_{m=k}^\infty \tilde{b}_m e^{-mx/2} \sqrt{\frac{m!}{(m-k)!}} |m-k\rangle 
\nonumber\\ & = & 
\sum_{m=0}^\infty \tilde{b}_{m+k} 
e^{-(m+k)x/2} \sqrt{\frac{(m+k)!}{m!}} |m\rangle, 
\end{eqnarray}
and thus 
\begin{eqnarray}
\label{eq:B_4}
\frac{\langle\Psi_\beta| \hat{a}^{\dagger \, k} \hat{a}^k 
|\Psi_\beta\rangle
}{N^k k!} & = & 
\frac{e^{-kx}}{N^k k!} \sum_{m=0}^\infty |\tilde{b}_{m+k}|^2 
\frac{(m+k)!}{m!} e^{-mx}
\nonumber\\ & \le & 
\frac{|\tilde{b}_{\rm max}|^2 e^{-kx}}{N^k k!} \frac{k!}{(1-e^{-x})^{k+1}}
\nonumber\\ & = & \frac{1}{N^k} 
\frac{|\tilde{b}_{\rm max}|^2}{1-e^{-x}} 
\left( \frac{e^{-x}}{1-e^{-x}} \right)^k .
\end{eqnarray}
This bound depends on $\tilde{b}_{\rm max}$ and $x$ i.e. 
the state $|\Psi_{\beta_i}\rangle$. Therefore, maximizing 
the rhs of this inequality for all $|\Psi_{\beta_i}\rangle$ and 
denoting the maximum value as $C_k^{\rm max}/N^k$,  
one obtains Eq.~(\ref{eq:B_1}).

\section{Total photon number statistics}
The exponential and Poissonian distributions are described as
\begin{equation}
\label{eq:C_1}
P_E (m) = C_E e^{-mx}, 
\end{equation}
and 
\begin{equation}
\label{eq:C_2}
P_P (m) = \frac{C_P^m}{m!}e^{-C_P}, 
\end{equation}
respectively. 
Then the distribution of the total photon number is given by 
\begin{eqnarray}
\label{eq:C_3}
P_{\rm tot} (n) & = & 
\sum_{m=0}^n P_P (m) P_E (n-m) 
\nonumber\\ & = & 
e^{-C_P} \sum_{m=0}^n \frac{C_P^m}{m!} C_E e^{-(n-m)x} 
\nonumber\\ & = & 
C_E e^{-(C_P + nx)} \sum_{m=0}^n \frac{(C_P e^x)^m}{m!} 
\nonumber\\ & \le & 
C_E e^{C_P (e^x-1)} e^{-nx}, 
\end{eqnarray}
which decreases exponentially as $n$ increases.


\begin{references}

\bibitem{KLM01}
   E.~Knill, R.~Laflamme, and G.~J.~Milburn,
   Nature\,\textbf{409}, 46 (2001).

\bibitem{vanLoock03}
   P.~van Loock and N.~L\"{u}tkenhaus,
   Phys.\ Rev.\ A\,\textbf{69}, 012302 (2004).

\bibitem{comment1}
This is true only for the case when all physical operations 
during a whole measurement can be described by rank 1 operators. 
As will be shown in the text, our scheme corresponds to this case. 

\bibitem{Walgate00}
   J.~Walgate, A.~J.~Short, L.~Hardy, and V.~Vedral, 
   Phys.\ Rev.\ Lett.\,\textbf{85}, 4972 (2000). 

\bibitem{vanLoock00}
   P.~van~Loock and S.~L.~Braunstein, 
   Phys.\ Rev.\ Lett.\,\textbf{84}, 3482 (2000). 

\bibitem{Barnett97}
$e^{\theta_i 
(\hat{a}^{\dagger}_i \hat{a}_0 
- \hat{a}_i \hat{a}^{\dagger}_0 )} = 
e^{-\hat{a}^{\dagger}_i \hat{a}_0 \tan\theta_i} 
e^{-\ln \cos\theta_i 
(\hat{a}^{\dagger}_i \hat{a}_i - \hat{a}^{\dagger}_0 \hat{a}_0 )} 
e^{\hat{a}_i \hat{a}^{\dagger}_0 \tan\theta_i} $
   has been used. See 
   S.~M.~Barnett and P.~M.~Radmore, 
   {\it Methods in Theoretical Quantum Optics} 
   (Oxford University Press, New York, 1997), for example.

\bibitem{Dolinar73}
   S. J. Dolinar,
   RLE, MIT, QPR No.~111, 1973 (unpublished) p.~115. 

\bibitem{Geremia04}
   J. M. Geremia,
   Phys.\ Rev.\ A\,\textbf{70}, 062303 (2004).

\bibitem{Takeoka05}
   M.~Takeoka, M.~Sasaki, P.~van Loock, and N.~L\"{u}tkenhaus, 
   Phys.\ Rev.\ A\,\textbf{71}, 022318 (2005).

\bibitem{C_k^max}
One can show that $C_k^{\rm max}$ takes a finite value, 
by using the exponential decay of the number distribution 
of $|\Psi_\beta\rangle$ and the relation 
$\sum_{m=0}^{\infty} z^m (m+k)!/m! = k!/(1-z)^{k+1}$. 
(see Appendix B for details). 


\bibitem{Cahill69}
   K.~E.~Cahill and R.~J.~Glauber, 
   Phys.\ Rev.\ \textbf{177}, 1857 (1969). 

\bibitem{HandbookMath}
{\it Handbook of Mathematical Functions}, ed. M.~Abramowitz 
and I.~A.~Stegun (Dover Publications Inc., New York, 1965).

\bibitem{GenIneq}
{\it General Inequalities 1}, ed. E.~F.~Beckenbach 
(Birkh\"{a}user Verlag, Basel, 1978).   


\end{references}
\end{document}